# Interpreting encoding and decoding models

Nikolaus Kriegeskorte[a] & Pamela K. Douglas[b]

[a] Department of Psychology, Department of Neuroscience, Department of Electrical Engineering, Zuckerman Mind Brain Behavior Institute, Columbia University, New York, NY

[b] Center for Cognitive Neuroscience, University of California, Los Angeles, CA; Modeling, Simulation, Computer Science, UCF, USA

**Abstract**

Encoding and decoding models are widely used in systems, cognitive, and computational neuroscience to make sense of brain-activity data. However, the interpretation of their results requires care. Decoding models can help reveal whether particular information is present in a brain region in a format the decoder can exploit. Encoding models make comprehensive predictions about representational spaces. In the context of sensory experiments, where stimuli are experimentally controlled, encoding models enable us to test and compare brain-computational theories. Encoding and decoding models typically include fitted linear-model components. Sometimes the weights of the fitted linear combinations are interpreted as reflecting, in an encoding model, the contribution of different sensory features to the representation or, in a decoding model, the contribution of different measured brain responses to a decoded feature. Such interpretations can be problematic when the predictor variables or their noise components are correlated and when priors (or penalties) are used to regularize the fit. Encoding and decoding models are evaluated in terms of their generalization performance. The correct interpretation depends on the level of generalization a model achieves (e.g. to new response measurements for the same stimuli, to new stimuli from the same population, or to stimuli from a different population). Significant decoding or encoding performance of a single model (at whatever level of generality) does not provide strong constraints for theory. Many models must be tested and inferentially compared for analyses to drive theoretical progress.

**Highlights**

1. Decoding models can reveal whether particular information is present in a brain region in a format the decoder can exploit.
2. Encoding models make comprehensive predictions about representational spaces and more strongly constrain computational theory.
3. The weights of the fitted linear combinations used in encoding and decoding models are not straightforward to interpret.
4. Interpretation of encoding and decoding models critically depends on the level of generalization achieved.

5. Many models must be tested and inferentially compared for analyses to drive theoretical progress.

## Encoding and decoding: concepts with caveats

The notions of encoding and decoding are rooted in the view that the brain processes information. Information about the world enters our brains through the senses, is processed and potentially stored, and informs our behavior over a large range of time scales. To understand brain function, then, we must understand the information processing: *what information* is processed, *in what format* it is coded in neural activity, and *how it is re-coded* across stages of processing, so as to ultimately contribute to behavior.

The view of the brain as an information-processing device (or, equivalently, a computational device) is closely linked to the concept of *representation*. A pattern of neural activity *represents* information about the world if it *serves the function* to convey this information to downstream regions, which use it to produce successful behavior (DeCharms & Zador 2000).

When we talk about encoding and decoding, we focus on a particular brain region X whose function we are trying to understand. To simplify the problem, we divide the process into the encoder (which produces the code in region X from the sensory signals) and the decoder (which uses the code in region X to enable successful behavior). This bipartite division often provides a useful simplification. However, we have to consider three caveats to avoid confusion.

First, encoder and decoder do not inherently differ with respect to the nature of the processing. Where the encoder ends and the decoder begins depends on our region of interest X. When we move our focus to the next stage of representation in region Y, the processing between X and Y, which was part of the decoder, becomes part of the encoder. Whether a particular processing step in the brain is to be considered encoding or decoding, thus, is in the eye of the beholder.

Second, our region of interest X may not be part of all causal paths from input to output. The division about region X into encoder and decoder, then, misses a portion of the causal graph. Brain regions do not in general form chains of processing stages without skipping connections or recurrent signaling. The primate visual hierarchy is a case in point, where cortical areas interact in a network with about a third of all possible pairwise inter-area connections (Felleman & Van Essen 1991, Young 1992). Encoding and decoding models are nevertheless useful in this context, providing a partial view of the encoded information and its format.

Third, the terms encoding and decoding suggest that each is the inverse of the other: The encoder maps stimuli to brain responses; the decoder maps brain responses to stimuli. However, if we would like to interpret our encoding and decoding models as models of brain computation, then both encoder and decoder ought to operate in the causal direction. The decoder should not map *back* to stimuli, but *on* to motor representations. An encoding model might predict neural activity elicited by images (Wu et al. 2006, Kay et al. 2008, Wen et al. 2018). A decoding model might attempt to predict category labels from neural activity, thus explicating information that might be reflected in a behavioral response (Majaj et al. 2015).

The notion that encoding is followed by its inverse begs the question why the original input is not simply copied. A possible motivation for a sequence of an encoder and its inverse is to temporarily compress the information, e.g., for passage through a bottleneck such as the optic nerve. A more general notion is that information is *re-coded* through a sequence of stages, possibly compressing it for more efficient representation (Simoncelli & Olshausen 2001), but also discarding unneeded information and converting needed information into a format that enables it to be exploited to control behavior (Tishby et al. 2000, Achille & Soatto 2018). When decoding models are conceptualized as inverse encoders (mapping back to the stimuli, rather than on to downstream representations), they cannot be interpreted as process models of brain function. However, they can still be useful tools, revealing information present in a brain region and giving us clues about the format of the code. For a careful discussion of the interpretation of causal and anti-causal encoding and decoding models, see Weichwald et al. (2015).

Encoding and decoding are important concepts in both theoretical and experimental neuroscience. This paper addresses experimentalists and focuses on the interpretation of empirical results obtained by fitting encoding and decoding models to brain-activity data.

## Decoding models: revealing information and its format

Decoding is sometimes described as "reading the brain" or "cracking the neural code" (Cox & Savoy 2003, Haynes & Rees 2006, Tong & Pratte 2012). The underlying concept is that of decoding as inverse encoding, where the goal is not to model brain information processing, but to reveal the content of the code. The sense of revealing a mystery and gaining insight into hidden information drives the imagination of scientists and lay people alike.

The intuition that something important has been learned when we can decode some perceptual or higher cognitive content from brain activity is correct. However, "decrypting" the brain should not be equated with understanding its computational mechanism. Decoding reveals the products, not the process of brain computation (Box 1). However, it is a useful tool for testing whether a brain region contains a particular kind of information in a particular format (Haxby et al. 2001, Cox & Savoy 2003, Carlson et al. 2003, Kamitani & Tong 2005, Haynes & Rees 2006, Kriegeskorte et al. 2006, Norman et al. 2006, Paninski et al. 2007, Mur et al. 2009, Pereira et al. 2009, Pillow et al. 2011, Kriegeskorte 2011, Tong & Pratte 2012, Hebart et al. 2015, Haynes 2015, Hebart & Baker 2017, Varoquaux et al. 2017, Paninski & Cunningham 2017).

A simple example of a decoding model is a linear classifier (Duda et al. 2012) that takes a measured brain-activity pattern as input and outputs a class label (say "cat" and "dog"), revealing which of two stimuli (say a particular image of a cat and a particular image of a dog) elicited the response pattern (Figure 1). A linear classifier may achieve this by computing a weighted sum of the measured responses and imposing a threshold to decide which label to return (see Mur et al. 2009 for an introduction in the context of neuroimaging). The weights are set to maximize the accuracy of the decoding on a training data set. If decoding succeeds reliably on a test data set, then the region must contain information about the decoded variable (the binary stimulus distinction here).

### Decoding provides a statistically advantageous way of testing for stimulus information

In the two-stimulus example, all the linear decoder shows is that the two images elicit distinct response patterns. This means that there is mutual information between stimulus and response. To demonstrate mutual information, we could have used an encoding model (Naselaris et al. 2011) or a multivariate test of difference between the response patterns elicited by the two stimuli (e.g. multivariate analysis of variance), instead of a decoding model. However, a univariate encoding model would in general have less sensitivity, because it does not account for the noise correlations between different response channels (Averbeck et al. 2006). A multivariate analysis of variance would account for the noise correlations, but might have less specificity. In fact, it might fail to control false-positives at the nominal level if its assumptions of multivariate normality did not hold (as is often the case), making it invalid as a statistical test (Kriegeskorte 2011). Decoding provides a natural approach to modelling the noise correlations (e.g. using a multivariate normal model as in the Fisher linear discriminant), without relying on the model assumptions for the validity of the test: Violations of the decoding model assumptions will hurt decoding performance. We, thus, err on the safe side of concluding that there is no information about the stimuli in the responses. In sum, it is not the direction of the decoding model ("brain reading") that makes it a compelling test for information, but the statistical nature of the problem (noise correlations) and the fact that decoders are tested on independent data.

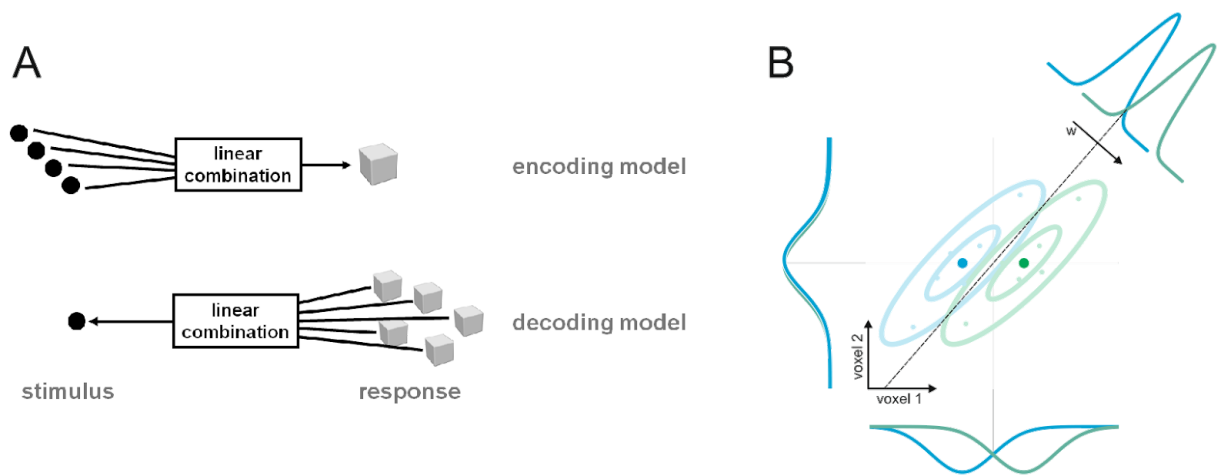

**Figure 1 | Linear encoding and decoding models. (A)** Encoding and decoding model the relationship between stimuli and responses in opposite directions. An encoding model (top) predicts brain responses as a linear combination of stimulus properties (black circles). A decoding model (bottom) predicts stimulus properties as a linear combination of brain responses. **(B)** Example of a linear decoding model using a 2-dimensional feature space consisting of two voxels. Voxel 1 contains relevant information about which of two classes (green, blue) the stimulus belongs to. Voxel 2 contains no information about the stimulus class. The two dimensions jointly define the linear discriminatory boundary. Note that the weights assigned to each voxel are defined by the vector w, which is orthogonal to the decision boundary. Because the noise is correlated between the voxels, a linear decoder will assign significant negative weight to voxel 2, using this voxel (which contains only noise) to cancel the noise in the voxel which contains signal. As a result, interpreting the absolute weights of linear decoders requires care and additional analyses.

## Linear decodability indicates "explicit" information

For decoding to succeed, the information must be present in the brain region in a format that the decoder can exploit. Linear decoders, the most widely used class, require that the distributions of patterns be linearly separable to some extent. This is a weakness in that we might fail to detect information encoded in a more complex format. However, it is a strength in that it provides clues to the format of the information we do detect. The simpler the decoder, the more focused its sensitivity will be. From the perspective of understanding the brain computations, it is attractive to use decoding operations that single neurons can implement. These include linear readout, but also simple nonlinear forms of readout such as radial basis function decoding (Poggio & Girosi 1990). Linear decodability indicates that a downstream neuron receiving input from a sufficient portion of the pattern, could read out the information in question (DiCarlo & Cox 2007). Information amenable to direct readout by downstream neurons is sometimes referred to as "explicit" in the code (Kriegeskorte 2011, DiCarlo et al. 2012, Hong et al. 2016).

## The level of generalization beyond the training set must be considered when interpreting a decoding result

Fitting a model always poses the risk of overfitting, i.e. of optimizing the fit to the training data at the expense of predictive performance on independent data. Overfitting can lead to high decoding accuracy on the training set, even if the response patterns contain no information about the stimulus. Decoders therefore need to be tested for generalization to independent data (Hastie et al. 2009, Kriegeskorte 2015). In our example, we might test the decoder on an independent set of response measurements for the same two particular images of a cat and a dog. If decoding accuracy on this independent test set is significant, we can reject the null hypothesis that the response patterns contain no information about the stimulus (Mur et al. 2009, Pereira et al. 2009).

However, detecting information about which of two images has been presented tells us almost nothing about the nature of the code. The two images must have distinct response patterns in the retina and V1 (low-level representations) as well as in the visuo-semantic regions of the ventral stream (high-level representations). We would therefore expect a linear decoder to work on new measurements for the same images in any of these regions. This reflects the fact that all the regions contain image information. In the retina, for example, we expect the two images to elicit distinct response patterns, while the manifolds of response patterns corresponding to the two categories are hopelessly entangled (DiCarlo & Cox 2007, Chung et al. 2016, Chung et al. 2018).

Given responses to just two images, we can demonstrate the presence of information, but have no empirical basis for characterizing the nature of the code (see Kriegeskorte et al. 2007 for a study limited by this drawback). In order to learn whether the region we are decoding from contains a low-level image encoding or a high-level categorical encoding, we can train the decoder on one set of cat and dog images and test it on another set of images of different cats and dogs (Anzellotti et al. 2014, Freedman et al. 2002, Kriegeskorte 2011).

To support the interpretation that "cats" and "dogs" are linearly separable in the representation (rather than the weaker claim that there is image information), it is not sufficient to increase the number of particular images of cats and dogs, while training and testing on the same images. The linear decoder has many parameters (one for each response channel) and is expected to overfit even to a larger set of particular images. Even for the retinal representation, we therefore expect a cat/dog decoder to generalize to new measurements performed on the same images. We must test the linear decoder for generalization to different cats and dogs.

Note, however, that interpreting linear decodability as linear separability of the two classes in the neuronal representational space would further require the decoding accuracy to be so high that errors can be attributed to the measurement noise rather than the neural representation. In practice, we typically face ambiguity. For example, decoding accuracy may be significantly above chance, but far from perfect. This indicates that the code contains some linearly decodable information, but claims of linear separability may be difficult to evaluate as it would require attributing the substantial proportion of errors to limitations of the measurements (noise and subsampling), rather than to a lack of linear separability of the neuronal activity patterns.

If we managed to show that cats and dogs fall in linearly separable regions in representational space, there would still be a question about the nature of the features that support the encoding. There may be many different visual features that can be computed from images and that lend themselves to separating cats from dogs (e.g. the shape of the ears or the shape of the eyes). Revealing which particular features the encoding in a given brain region employs would require further experiments. For example, we could design artificial stimuli that contain subsets of the distinctive features of cats and dogs. We could then test whether a linear decoder trained on responses to natural images of cats and dogs generalizes to responses to the artificial stimuli. Successful decoding would support the hypothesis that the brain region employs some of the features present in the artificial images.

More generally, we can go beyond using different exemplars from the same categories in the test set. Testing decoders for cross-generalization between different domains is useful to address a host of questions about the invariances of the encoding, for example its consistency between imagery and perception (Stokes et al. 2009, Cichy et al. 2012) and its stability across latencies after stimulus onset (King & Dehaene 2014).

## Stimulus reconstruction provides a richer view of the information present, but complex priors complicate the interpretation

A decoder may discriminate more than two categories. It might decode a continuous variable (e.g. the orientation of the stimulus; Kamitani & Tong 2005), or it might output a multidimensional description of the stimulus. The most ambitious variant of decoding is stimulus reconstruction, where the output is a detailed estimated rendering of the stimulus. A decoder providing a rendering of the stimulus can potentially reveal more of the information present in the neuronal encoding (Thirion et al. 2006, Miyawaki et al. 2008, Mesgarani et al. 2009, Naselaris et al. 2009, Santoro et al. 2017, Parthasarathy et al. 2017).

Stimulus reconstruction is a particularly impressive feat of decoding for two related reasons. First, the space of outputs is much richer, so the decoder reveals more of the encoded information (discriminating among more possible stimuli). Second, a more challenging level of generalization is typically required for the decoder to discriminate among this richer set of possible stimuli, because there are usually severe limitations on the number of stimuli that the decoder can be directly trained on. Successful generalization suggests that the structure of the model captures something about the code that enables generalization far beyond the training set.

To the extent that stimulus reconstruction, when applied to novel stimuli that the model has not been trained on, successfully specifies close matches within a rich space of reconstruction targets (e.g. pixel images), the analysis indicates that the code contains rich information about the stimulus. An extreme example of this would be a decoder that can precisely reconstruct arbitrary natural or unnatural images from their representation in V1. This would indicate that V1 encodes the image precisely and that the encoding is not restricted to natural images. In fact, this has never been shown for V1 and would be puzzling, because we expect visual representations to be specialized for natural stimuli.

**Box 1: Decoding Models: Benefits and Limitations**

Decoding provides an intuitive and compelling demonstration of the presence of information in a brain region. Decoding models bring several benefits:

- They enable researchers to look into the regions and reveal whether particular information is encoded in the responses. Encoded information might be used by downstream neurons (i.e. the encoding might serve as a representation). Training and testing decoders has helped move the focus from single-neuron activity and regional-mean activation to the information being processed, making analyses more relevant to the computational function in question.

- They enable cognitive neuroscience to exploit the fine-grained multivariate information present in modern imaging data. Before the advent of decoding, dominant brain mapping techniques required smoothing of the data, treating fine-grained information as noise. The multivariate modelling of local patterns also brought locally multivariate noise models to the literature, which greatly enhance the information that can be drawn from neuroimaging data.

- Decoding analyses use independent test data to assess the presence of information. Demonstrating significant information with independent test data ensures that violations of model assumptions lead to conservative errors (making significant results less likely). This is an advantage over methods (such as univariate and multivariate linear regression) that use distributional assumptions rather than independent test sets to perform statistical inference.

Decoding models are limited in several ways:

- They cannot in general be interpreted as models of brain-information processing. In the context of sensory systems, decoding models operate in the wrong direction, capturing the inverse of the brain's processing of the stimuli. If the decoded variable is a behavioral response, then direction of information flow matches between model and brain. However, such applications are rare. Moreover, the most successful applications of decoding in the literature rely on linear decoders, which are useful to reveal explicit information, but cannot capture the complex nonlinear computations we wish to understand.

- Decoder weight maps are difficult to interpret (Figure 3) because voxels uninformative by themselves can receive large weights when they help cancel noise and because the weights are codetermined by the data and the prior, and the fitting procedure might select one but not the other of two essentially equally informative voxels.

- Decoding results provide only weak constraints on computational mechanisms. Knowing that a particular kind of information is present provides some evidence for and against alternative theories about what the region does. However, it does not exhaustively characterize the representation, let alone reveal the underlying computations.

In order to improve the appearance of reconstructions of natural stimuli, it is attractive to use a prior over the possible outputs of the decoder. For example, we might constrain the model to output an image whose low-level or high-level statistics match natural images. We may constrain the decoder more strongly, by limiting it to output only actual natural images. This constraint could be implemented using a restricted set of target images (Naselaris et al. 2009) or a generative model of natural images (Goodfellow et al. 2014). In either case, the quality of the reconstruction will reflect a combination of the information provided by the brain region and the information contributed by the prior. A complicated prior therefore also complicates the interpretation of the reconstruction results: good looking reconstructions no longer indicate that all the detail they provide is encoded in the brain region. The reconstruction has to be compared to the presented stimulus, and the complexity of the output space (which is reduced by the prior over the outputs) needs to be considered in the interpretation.

An important question is what we can learn from stimulus reconstructions. The goal to learn about the content and format of the code may not be ideally served by striving for the most natural looking reconstruction.

### Decoding models predicting behavioral responses from brain responses can be interpreted as brain-computational models

Decoding is usually used as a tool of analysis that reveals aspects of the content and format of the information encoded in a brain region. The decoding model, thus, is not interpreted as a model of brain computation. In the context of sensory systems, a decoder maps from brain responses to stimuli. Since stimulus processing by the brain operates in the opposite direction, it is difficult to interpret a decoder as a model of brain information processing. However, if a decoder is used to predict behavioral responses, e.g. judgments of categorical or continuous stimulus variables (possibly including errors and reaction times on individual trials), then the decoder can be interpreted as a model (at a high level of abstraction) of the brain computations generating the behavioral responses from the encoding of the stimuli in the decoded brain region (Shadlen et al. 1996, Williams et al. 2007, Walther et al. 2012).

## Encoding models: testing comprehensive representational predictions

Encoding models attempt to predict brain response patterns from descriptions of the experimental conditions (Figure 1A; Paninski et al. 2007, Kay et al. 2008, Dumoulin & Wandell 2008, Mitchell et al. 2008, Naselaris et al. 2011, Khaligh-Razavi & Kriegeskorte 2014, Yamins et al. 2014, Naselaris & Kay 2015, van Gerven 2017). Encoding models, thus, operate in the opposite direction as decoding models.

If our goal is merely to demonstrate that a brain region contains information about the experimental conditions, then the direction the model should operate in is a technical issue: One direction may be more convenient for capturing the relevant statistical dependencies (e.g. noise correlations among responses), but a model operating in either direction could support the inference that particular information is present in the code. If our goal is to test computational theories, however, then the direction that the model operates in matters, because it determines whether the model can be interpreted as a brain-computational model.

### Encoding models predicting brain responses from sensory stimuli can serve as brain-computational models

Whereas a decoding model typically serves to test for the presence of particular information in a brain region, an encoding model can provide a process model, at some level of abstraction, of the brain computations that produce the neuronal code. An encoding model makes comprehensive predictions about the representational space (Naselaris & Kay 2015, Diedrichsen & Kriegeskorte 2017; Box 2) and, ideally, should fully explain neuronal responses in the region in question (up to the maximum achievable given the noise in the data).

In sensory neuroscience, the experimental conditions correspond to sensory stimuli, and so an encoding model maps sensory stimuli to their encodings in sensory regions of the brain (Kay et al. 2008, Mitchell et al. 2008, Dumoulin & Wandell 2008, Huth et al. 2012). A sensory encoding model, thus, operates in the direction of the information flow: from stimuli to brain responses. It should take raw sensory stimuli as input (e.g. images or sounds; Kay et al. 2008, Naselaris et al. 2011, Khaligh-Razavi & Kriegeskorte 2014, Yamins et al. 2014, Kell et al. 2018; for recent reviews see Kriegeskorte 2015, Yamins & DiCarlo 2016) and predict the patterns of brain activity the stimuli elicit. We can adjudicate among computational theories by implementing them in encoding models and testing their ability to predict brain-activity patterns (Kriegeskorte & Diedrichsen 2016).

## Brain-computational encoding models can be tested by predicting raw measurements, representational dissimilarities, or the activity-profiles distribution

A brain-computational encoding model could consist in a set of hand-engineered features computed from the stimuli or in a neural network model trained on some task. How can we evaluate a high-dimensional representation in a brain-computational encoding model with brain-activity data, when the units of the model may not correspond one-to-one to the measured channels of brain activity?

One approach is to predict the raw measurements (Kay et al. 2008, Dumoulin & Wandell et al. 2008, Mitchell et al. 2008, Naselaris et al 2011, van Gerven 2017). To achieve this, we can fit a linear model that maps from the outputs of the nonlinear encoder thought to capture the brain computations (e.g. a layer of a neural network model processing the stimuli) to each measured response channel, e.g. each neuron or fMRI voxel in the region we would like to understand. The linear model will capture which units of the nonlinear encoding model best predict each response channel. It can be interpreted as capturing the sampling and averaging that occurs in the measurement of brain activity. A neuronal recording array, for example, will capture a sample of neurons, and fMRI will give us measurements akin to local spatiotemporal averages at regular grid locations. For each response channel, the linear model will have a parameter for each nonlinear feature (e.g. each unit of the neural network layer). Fitting these parameters requires substantial training data and normally also a prior on the parameters (e.g. a 0-mean Gaussian prior as is implicit to fitting with an L2 penalty).

Instead of predicting the raw measurements, we can use a brain-computational encoding model to predict to what extent different stimuli are dissimilar in the representation, an approach called representational similarity analysis (RSA; Kriegeskorte et al. 2008, Nili et al. 2014). The pairwise dissimilarities of the multivariate response patterns representing the stimuli can be summarized in a representational dissimilarity matrix (RDM). For example, for each pair of stimuli, the dissimilarity between the associated response patterns could be measured using the Euclidean distance. The RDM characterizes the geometry of the set of points in the multivariate response space that correspond to the stimuli. Noise correlations can be captured by estimating the covariance matrix of the residuals of the response-estimation model, and replacing the Euclidean distance by the Mahalanobis distance. To remove the positive bias associated with measuring distances between noisy data points, we can use the crossnobis (crossvalidated Mahalanobis) estimator (Nili et al. 2014, Walther et al. 2016). The resulting crossnobis RDM provides a full characterization of the linearly decodable information in the representational space (Diedrichsen & Kriegeskorte 2017). Comparing representations in models and brains at the level of RDMs obviates the need for fitting a linear model to predict each measured response (thus reducing the need for training data) and enables the analysis to naturally handle noise correlations between responses (which are typically ignored when encoding models separately predict each of the measured response channels).

A third approach to the evaluation of encoding models is to predict the distribution of activity profiles. In pattern component modelling (PCM, Diedrichsen et al. 2011), this distribution is characterized by the second moment of the activity profiles. Like the RDM, this is a stimulus-by-stimulus summary statistic of the stimulus-response matrix. Each entry of the second-moment matrix corresponds to the inner product between two response patterns.

All three approaches can be construed as testing hypotheses about the representational space induced by the activity profiles (Diedrichsen & Kriegeskorte 2017). Consider a linear encoding model using a Gaussian prior on the weights. Such a model predicts a Gaussian distribution of activity profiles. The predicted distribution of activity profiles is captured by its second moment. For representational similarity analysis, similarly, the RDM is a function of the second moment of the activity profiles. This core mathematical commonality between the methods notwithstanding, each is best suited for a particular set of questions.

Linear models predicting raw measurements lend themselves to univariate brain mapping, revealing which voxels or neurons are accounted for by a particular nonlinear encoding model. RSA lends itself to characterizing the geometry of the representational space, naturally handles noise correlations among responses, and reduces the need for training data. PCM can have greater sensitivity for adjudicating among models than the other two methods, at the expense of relying on stronger assumptions. The three methods are best viewed as part of a single toolbox of representational model analyses, whose elements can be combined as needed to address particular questions.

### The level of generalization beyond the training set must be considered when interpreting an encoding result

Encoders, like decoders, are tested by evaluating how well they predict independent data, whether the predicted quantities are the raw brain-activity measurements, the representational dissimilarity matrix, or the second-moment matrix of the activity profiles. For encoders, as for decoders, the interpretation depends on both the prediction accuracy and the level of generalization beyond the training set that the model achieves.

Encoding models typically require the fitting of parameters, so overfitting needs to be accounted for in any inferential procedure. In the simplest type of a univariate linear encoding model, we can rely on Gaussian assumptions and perform inference without a separate test set (e.g. Friston et al. 1994). However, more interesting models require independent test sets, for example when parameters are fitted using priors over the weights and when the model is a brain-computational model to be tested for generalization to new conditions.

A key consideration is how much flexibility to allow in fitting each model representation to a brain representation. One extreme is to allow no flexibility and assume that the model representation precisely predicts the geometry of the representational space (Kriegeskorte et al. 2008). This case is most naturally handled by RSA and PCM, but could also be implemented with linear encoders by using a prior that prevents any distortion of the representational geometry. The other extreme is to allow arbitrary linear remixing of the units of the nonlinear encoding model. This case is most naturally handled with linear encoding models, but can also be implemented with PCM and RSA (Diedrichsen et al. 2011, Khaligh-Razavi & Kriegeskorte 2014, Khaligh-Razavi et al. 2016). In practice, some compromise is desirable, which we can think of as a prior on the mapping from the brain-computational model to the measured brain responses. We might use a 0-mean Gaussian prior on the weights (e.g. Kay et al. 2008). Alternatively, we can limit flexibility more aggressively, by allowing each unit (or each feature map or layer) a single weight (not a separate weight for each response). Such weighted representational models (e.g. Khaligh-Razavi et al. 2014) are naturally implemented with RSA and PCM. Each brain-computational model in this case predicts a superset of the features spanning the brain representational space (disallowing linear mixtures), but does not predict the prevalence of each of the features in the neuronal population.

The lowest level of generalization beyond the training set is generalization to new measurements for the same experimental conditions. This is sufficient, if the experimental conditions exhaustively cover the domain we would like to draw inferences about (consider the case of the representation of the five fingers in motor cortex: Diedrichsen et al. 2012). However, in a domain such a sensory systems, the goal is typically to evaluate to what extent a brain-computational model can predict brain representations of arbitrary stimuli. This requires a higher level of generalization beyond the training set. A vision model, for example, might be trained with responses to one sample of natural images and tested for generalization to responses to an independent (and nonoverlapping) sample from the same distribution of natural images. Because the set of all natural images is so rich, this is a challenging generalization task (as illustrated by the difficulty of computer vision). An even more stringent test of the assumptions implicit to a model is to train the model on a sample from one population of images and test it on a sample from a different population of images (e.g. Eickenberg et al. 2017).

The prediction accuracy can be assessed in terms of whether it is significantly above chance level, whether it significantly differs from that of competing encoding models, and how close it comes come to the noise ceiling (the highest achievable accuracy given the noise in the data, Nili et al. 2014).

We can generalize claims about an encoding model to the extent that its predictions generalize. If we want to conclude that the model can predict responses for the stimuli presented, we need not test the model with different stimuli (only with different response measurements elicited by the same stimuli). If we want to conclude that the model can predict responses to arbitrary natural stimuli, we need to test it with new arbitrary natural stimuli. The population of conditions the test set is a sample of defines the scope of the claims we can make (Hastie et al. 2009, Kriegeskorte 2015).

We focus here on encoding and decoding models that are fitted to individual subjects' brains, so as be able to exploit fine-grained idiosyncratic patterns of activity that are unique to each subject. Within-subject prediction accuracy may support generalization to a population of stimuli, but it doesn't support generalization to the population of subjects. In some fields, such as low-level vision, researchers draw on prior knowledge and *assume* that results that hold for a few subjects will hold for the population. If we were instead to use our data to generalize our inferences to the population of subjects, we would either need to predict results for held-out subjects (which would make it impossible to exploit individually unique fine-grained activity patterns) or perform inference on the within-subject prediction accuracies with subject as a random effect.

## The feature fallacy: Interpreting the success of a model as confirmation of its basis set of features

Linear encoding models predict each measured activity profile as a linear combination of a set of model features. When a model can explain the measured activity profiles, we might be tempted to conclude that the model features are encoded in the measured responses. This interpretation is problematic because the same linear space can be spanned by many alternative basis sets of model features.

The fact that multiple sets of basis vectors can span the same space is widely appreciated. However, it is not obvious whether the ambiguity is removed when a prior over the encoding weights is used. An encoding model with a 0-mean isotropic Gaussian prior on the weights (equivalent to ridge regression) predicts a Gaussian distribution of activity profiles (captured by the second moment of the activity profiles as the sufficient statistic) and a particular representational geometry (captured by the RDM). The use of a weight prior does reduce the ambiguity. In the absence of a weight prior, all models spanning the same space of activity profiles make equivalent predictions. With a weight prior, different models spanning the same space can predict distinct distributions of activity profiles. However, substantial ambiguity remains (Figure 2) because there are infinitely many alternative ways to express the same distribution of activity profiles by different feature sets (each assuming a 0-mean isotropic Gaussian prior over the weights).

The freedom to choose among feature sets generating the same distribution of activity profiles is useful in that it enables us to implement a nonisotropic Gaussian prior on the weights of a given model (Tikhonov regularization) by re-expressing the model such that the features induce the same distribution of activity profiles in combination with a 0-mean isotropic Gaussian prior (Nunez-Elizalde et al. 2018). However, this freedom to express the same model by different feature sets needs to be kept in mind when interpreting results: Many alternative feature sets would have given the same results.

In sum, the same distribution of activity profiles can be expressed in many ways (by different feature sets), so the fact that a model explains brain responses does not provide evidence in favor of the particular feature set chosen. Diedrichsen (2018) termed the interpretation in terms of the particular features used the "feature fallacy". This fallacy is arguably somewhat persistent in the neuroscience literature (Churchland et al. 2012).

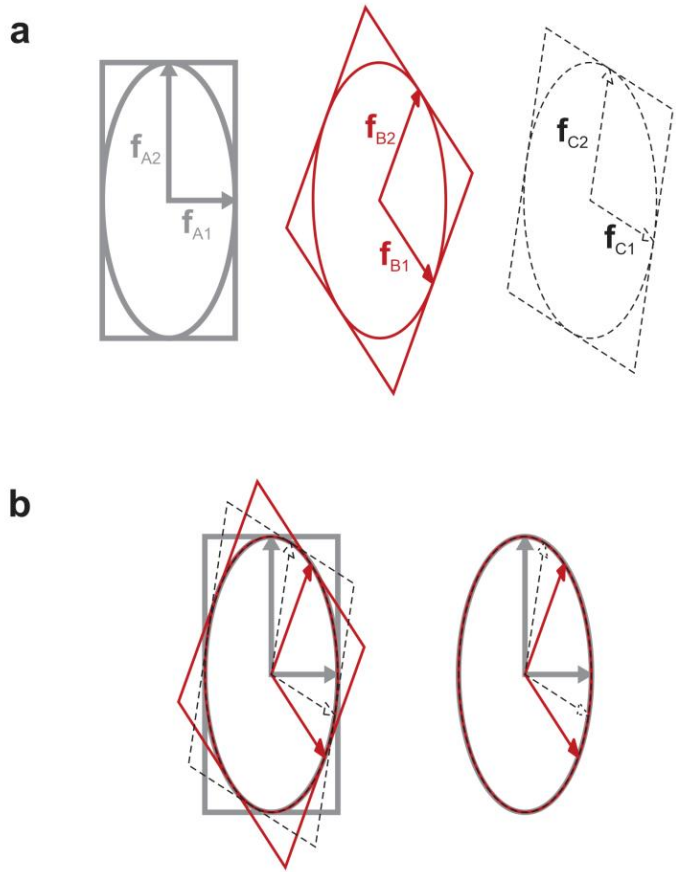

**Figure 2 | The feature fallacy.** Different linear encoding models spanning the same space of activity profiles may not be distinguishable. There are many alternative sets of feature vectors {$f_1$, $f_2$, …} that span the same space of activity profiles. In the absence of a prior on the weights of the linear model, all these sets can equally explain a given set of brain responses. The ambiguity is reduced, but not resolved when a prior on the weights is assumed (Diedrichsen et al. 2018). If we define a prior on the weights, then each model predicts a probability density over the space of activity profiles. This probabilistic prediction may be distinct for two sets of basis features, even if they span the same space. For example, if the weight prior is a 0-mean isotropic Gaussian, then each model assigns probabilities to different activity profiles according to a Gaussian distribution over the space of profiles. Two linear models may span the same space, but predict distinct distributions of activity profiles. However, even with a Gaussian weight prior, there are still (infinitely) many equivalent models that make identical probabilistic predictions. We illustrate this by example. (**a**) Three models (A, B, C) each contain two feature vectors as predictors (A: {$f_{A1}$, $f_{A2}$}, B: {$f_{B1}$, $f_{B2}$}, C: {$f_{C1}$, $f_{C2}$}). The three models all span the same 2-dimensional space of activity profiles. For each model, we assume a 0-mean isotropic Gaussian weight prior. (**b**) All three models predict the same nonisotropic Gaussian probability density over the space of activity profiles (indicated by a single iso-probability-density contour: the ellipse). Model A (gray) predicts the density by modeling it with two orthogonal features that capture the principal-component axes, with features having different norms to capture the anisotropy. Model B predicts the same density by modeling with two correlated features of similar norm. Model C falls somewhere in between, combining feature correlation and different feature norms to capture the same Gaussian density over the activity profiles. Note that there are many other models that span the same space, but will not induce the same probability density over activity profiles when complemented by a 0-mean isotropic Gaussian weight prior. A given linear encoding model's success at predicting brain responses provides evidence for the induced distribution of activity profiles, but not for the particular features chosen to express that distribution.

**Box 2: Encoding Models: Benefits and Limitations**

An encoding model predicts the response of each measurement channel (e.g. a neuron or fMRI voxel) on the basis of properties of the experimental condition (e.g. sensory stimulus). Continuous brain maps can be obtained by fitting such a model to each response in turn. Responses are typically predicted as linear combinations of the features, rendering this approach closely related to classical univariate brain mapping. However, encoding models have several important benefits:

- More complex feature sets are used, often comprising thousands of descriptive features. Fitting the models therefore requires regularization. Typically, models are fitted using a penalty on the weights.

- Encoding models can have nonlinear components, such as the location and size of a visual receptive field.

- Models are tested for generalization to different experimental conditions (e.g. a different sample of visual stimuli).

- When sensory systems are studied, encoding models operate in the direction of information flow. They can then be used as a general method for testing brain-computational models, i.e. models that process sensory stimuli. A vision model, for example, can take a novel image as input and predict responses to that image in cortex.

Analyses using encoding models to predict each response channel separately are limited in two ways:

- The separate modeling of each response channel is inconsistent with the multivariate nature of neuronal population codes and noise, and is statistically suboptimal. Separate modeling of each response entails low power for testing and comparing models, for two reasons: (1) Single responses (e.g. fMRI voxels) may be noisy and the evidence is not combined across locations. (2) The analyses treat the responses as independent, thus forgoing the benefit exploited by linear decoding approaches to model the noise multivariately. This is important in fMRI, where nearby voxels have highly correlated noise. As spatial resolution increases, we face the combined challenge of more and noisier voxels. This renders mapping with proper correction for multiple testing very difficult. In addition, relating results between individuals is not straightforward.

- When model parameters are color-coded and mapped across the cortex, the resulting detailed maps are not straightforward to interpret. (1) Weight maps of linear models are problematic to interpret in encoding models for the same reason they are problematic to interpret for decoding models: because a predictor's weight depends on the other predictors present in the model (unless each predictor is orthogonal to all others). The required regularization further complicates weight interpretation. (2) Models are fit at many locations and voxels highlighted on the basis of model fits. Post-selection inference on parameters is not usually performed. Because of these complications, results of fine-grained mapping across voxels are difficult to substantiate by formal inference.

Influential studies have met these challenges by focusing on single-subject analyses, acquiring a large amount of data in each subject, and limiting formal inference with correction for multiple testing to the overall explained variance of a model, with colors serving exploratory purposes.

# Weights of linear models are not straightforward to interpret in either encoding or decoding models

Beyond interpretation of the overall success of an encoding or decoding model, researchers often want to dig deeper and interpret the fitted parameters of their models. In the context of a decoding model, the weights assigned to the voxels would seem to tell us where the information the decoder uses is located in the brain. Similarly, in the context of an encoding model, the weights of different model features promise to reveal to what extent different model features are encoded in a brain region.

Unfortunately, the interpretation of the weights of linear models is not as straightforward as the simplicity of a linear relationship might suggest. A weight does not reflect the predictive power of an individual predictor (a measured brain response in decoding, a model features in encoding). Rather a weight reflects a predictor's contribution in the context of the rest of the model.

Uninformative predictors can receive large positive or negative weights. For example, an fMRI voxel that does not by itself contain any information can have a large absolute weight in a linear decoder if it improves decoding accuracy by cancelling noise that the voxel shares with other voxels that do contain stimulus information (Figure 1B; Figure 3; Haufe et al. 2014). In an encoding model, similarly, a model feature might contain no information about the modeled response and still receive a large absolute weight. For example, an fMRI voxel that only responds to faces might be explained by a set of semantic descriptors unrelated to faces if the nonface semantic descriptors are capable, in combination, of capturing contextual variation that is correlated with the presence of faces.

Conversely, informative predictors may receive weights that are small in absolute value or zero. For example, in fMRI, a voxel might receive zero weight when other voxels suffice for decoding when a weight penalty (especially a sparsity encouraging weight penalty, e.g. defined by the L1 norm of the weights) leads the fitting procedure to select among equally informative voxels (Figure 3). Weight penalties can similarly suppress model feature weights in encoding models.

A related problem is performing statistical inference to test hypotheses about the weights (Taylor & Tibshirani 2015). Systems and cognitive neuroscience has yet to develop a proper set of methods for hypothesis testing in this context.

A simple remedy to the complications associated with the interpretation of fitted weights is to interpret only the *accuracy* of decoding and encoding models (and its significance level) and not the parameters of the models. In the context of decoding, this makes sense for local regions of interest corresponding, for example, to cortical areas, which we can test one by one for the presence of particular information. It can be generalized to brain mapping by applying the decoder (or more generally any multivariate analysis) within a searchlight that scans the brain for the effect of interest (Kriegeskorte et al. 2006). Like classical univariate brain mapping, this approach derives its interpretability from the fact that each location is independently subjected to the same analysis. However, instead of averaging local responses, the evidence is summarized using local multivariate statistics. In the context of encoding models, similarly, we can focus on each model's overall performance and on inferential comparisons among multiple models.

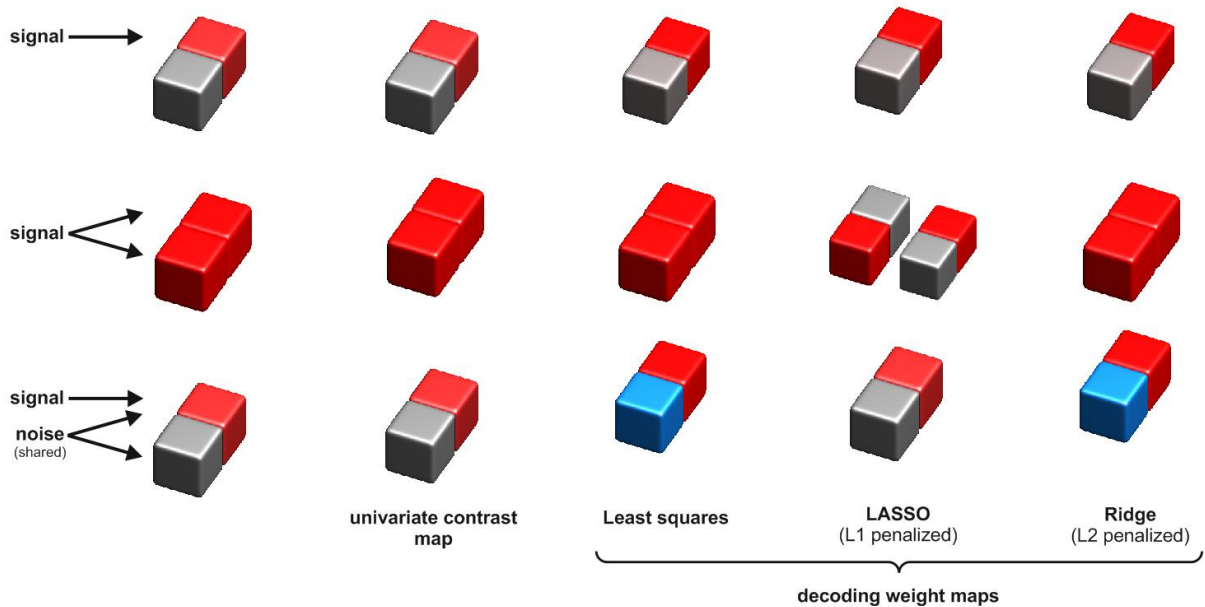

**Figure 3 | Weights of decoding and encoding models are difficult to interpret.** Three examples (rows) illustrate the difficulty of interpreting decoder weights for a pair of voxels. In the first example (top row), only the top voxel contains signal (stimulus information, red) and the two voxels have independent noise. This scenario is unproblematic: both univariate mapping (second column from the left) and decoder weight maps detect the informative voxel (red). In the second example (second row), both voxels contain the same signal. Here, univariate mapping and weight maps often work. However, the LASSO decoder, because of its preference for a sparse solution, may choose one of the voxels arbitrarily. In the third example (third row), only the top voxel contains signal and both voxels contain correlated noise. Univariate mapping correctly identifies the informative voxel. Linear decoders will give negative weight (blue) to the uninformative voxel, so as to cancel the noise.

# The single-model-significance fallacy

When our goal is merely to detect information in a brain region, we don't interpret the model as a model of brain computation. This lowers the requirements for the model: It need not operate in the direction of information flow and it need not be neurobiologically plausible. The model is merely a statistical tool to sense a dependency. The choice of model in this scenario, will affect our sensitivity, and its structure is not entirely irrelevant to the interpretation. For example, decodability by a linear model tells us something about the format of the encoding. However, a single model will suffice to demonstrate the presence of information in a brain region.

When our goal is to gain insight into the computations the brain performs, a model has a more prominent role: it is meant to capture, at some level of abstraction, the computations occurring in the brain. We then require the model to operate in the causal direction and to be neurobiologically plausible (albeit abstracted). Examples of such models include encoding models of sensory responses and decoding models that predict behavioral responses from brain activity. Psychophysical models, which skip the brain entirely and predict behavioral responses directly from stimuli, also fall into this class. In all these cases, finding that a model explains significant variance is a very low bar and tells us little as to whether the model captures the computational process.

The single-model-significance fallacy is to interpret the fact that a single model explains significant variance as evidence in favor of the model. A simple example that is widely understood is linear correlation. A significant linear correlation does demonstrate a dependency between two variables, but it does not demonstrate that the dependency is linear. Similarly, the fact that a complex encoding model explains significant variance in the responses of a brain region to a test set of novel stimuli does demonstrate that the brain region contains information about the stimuli, but it does not demonstrate that the encoding model captures the process that computes the encoding.

Even a bad model can explain significant variance, especially if it has a large number of parameters fitted to the data. In order to learn about the underlying brain computations, we need to (a) consider multiple models, (b) assess what proportion of the explainable (i.e. nonnoise) variance each explains at a given level of generalization, and (c) compare the models inferentially.

# Representational interpretations require additional assumptions

Decoding and encoding models are often motivated by the goal to understand how the brain represents the world, as well as the animal's decisions, goals, plans, actions, and motor dynamics. Significant variance explained by encoding and decoding models demonstrates the presence of information. Interpreting this information as a representation (Dennett 1987) implies the additional claim that the brain activity *serves the purpose* to convey the information to other parts of the brain (Kriegeskorte & Bandettini 2007, Kriegeskorte 2011, Diedrichsen & Kriegeskorte 2017). This functional interpretation is so compelling in the context of sensory systems that we sometimes jump too easily from findings of information to representational interpretations (de Wit et al. 2016, Ritchie et al. 2017). In addition to the presence of the information, its functional role as a representation implies that the information is read out by other regions, affecting downstream processing and ultimately behavior. Combining encoding and decoding models with stimulus- and response-based experimentation can help disambiguate the causal implications (Weichwald et al. 2015). Ideally, experimental control of neural activity should also be used to test whether activity has particular downstream or behavioral consequences (Afraz et al. 2006, Raizada & Kriegeskorte 2010). To the extent that we rely on prior assumptions to justify a representational interpretation, it is important to reflect on these and consider if there is evidence from previous studies to support them.

how these methods provide complementary tools in a single toolbox for representational analyses.